\documentclass[preprint,preprintnumbers,amsmath,amssymb]{revtex4}
\usepackage{amsfonts}    % list packages between braces
\usepackage{amssymb}
\usepackage{latexsym}
\usepackage{graphicx}
\usepackage[usenames,dvipsnames]{color}

\newcommand{\al}{\alpha}
\newcommand{\pa}{\partial}

\newcommand{\si}{\sigma}
\newcommand{\la}{\lambda}
\newcommand{\ta}{\tau}

\newcommand{\om}{\omega}

\newcommand{\ka}{\kappa}
\newcommand{\De}{\Delta}

\newcommand{\half}{\frac{1}{2}}
\newcommand{\rar}{\rightarrow}
\newcommand{\lrar}{\leftrightarrow}

\begin{document}

\title{Quasi-exact-solvability of the $A_{2}/G_2$ Elliptic model: algebraic forms, $sl(3)/g^{(2)}$ hidden algebra, polynomial eigenfunctions}

\author{Vladimir V~Sokolov\\
  Landau Institute for Theoretical Physics,
 Moscow Region, Russia\\
 and\\
Institut des Hautes Etudes Scientifique, Bures-sur-Yvette 91440, France\\
vsokolov@landau.ac.ru\\[12pt]
{\it and} \\[12pt]
Alexander V~Turbiner\\
Instituto de Ciencias Nucleares, UNAM, M\'exico DF 04510, Mexico\\
and\\
Institut des Hautes Etudes Scientifique, Bures-sur-Yvette 91440, France\\
turbiner@nucleares.unam.mx, turbiner@ihes.fr}

\begin{abstract}
The potential of the $A_2$ quantum elliptic model (3-body Calogero-Moser elliptic model) is defined by the pairwise three-body interaction through Weierstrass $\wp$-function and has a single coupling constant. A change of variables has been found, which are $A_2$ elliptic invariants, such that the potential becomes a rational function, while the flat space metric as well as its associated vector are polynomials in two variables. It is shown that the model possesses the hidden $sl(3)$ algebra - the Hamiltonian is an element of the universal enveloping algebra $U_{sl(3)}$ for arbitrary coupling constant - thus, it is equivalent to $sl(3)$-quantum Euler-Arnold top. The integral, in a form of the third order differential operator with polynomial, is constructed explicitly, being also an element of $U_{sl(3)}$. It is shown that there exists a discrete sequence of the coupling constants for which a finite number of polynomial eigenfunctions, up to a (non-singular) gauge factor occur. For these values of the coupling constant there exists a particular integral: it commutes with the Hamiltonian in action on the space of polynomial eigenfunctions, and the Hamiltonian is invariant with respect to two-dimensional projective transformations. It is shown that $A_2$ model has another hidden algebra $g^{(2)}$ introduced in Rosenbaum et al. \cite{Turbiner:1998}.

The potential of the $G_2$ quantum elliptic model (3-body Wolfes elliptic model) is defined by the pairwise and three-body interactions through Weierstrass $\wp$-function and has two coupling constants. A change of variables has been found, which are $G_2$ elliptic invariants, such that the potential becomes a rational function, while the flat space metric as well as its associated vector are polynomials in two variables. It is shown the model possesses the hidden $g^{(2)}$ algebra. It is shown that there exists a discrete family of the coupling constants for which a finite number of polynomial eigenfunctions up to a (non-singular) gauge factor occur. For these values of the coupling constants, there exists a particular integral and the Hamiltonian is invariant with respect to two-dimensional polynomial transformations.

\end{abstract}

\date{December 27, 2014}

%\preprint{IHES/P/14/29}

\vskip 2cm

\pacs{}

\maketitle

The $A_{2}$ elliptic model (3-body elliptic Calogero-Moser model, see e.g. \cite{Olshanetsky:1983}) describes three particles on the real line with pairwise interaction given by the Weierstrass $\wp$-function. It is characterized by the Hamiltonian
\begin{equation}
\label{OPHam}
 {\cal H}^{(e)}_{A_2}\ =\ -\ \half \sum_{i=1}^3\frac{\pa^{2}}{\pa x_i^{2}}\ +\
 \nu (\nu -1) \ \bigg(\wp (x_1-x_2)\ +\ \wp (x_2-x_3) \ +\ \wp (x_3-x_1)\bigg)\ \equiv\ - \half \De^{(3)} + V\ ,
\end{equation}
where $\De^{(3)}$ is three-dimensional Laplace operator, $\ka \equiv \nu(\nu-1)$ is coupling constant. The Weierstrass function $\wp (x) \equiv \wp (x|g_2,g_3)$ (see e.g. \cite{WW:1927}) is defined as
\begin{equation}
\label{wp}
    (\wp'(x))^2\ =\ 4\ \wp^3 (x) - g_2\ \wp (x)\ -\ g_3\ =\ 4(\wp (x) -e_1)(\wp (x) -e_2)(\wp (x) -e_3),
\end{equation}
where $g_{2,3}$ are its invariants and $e_{1,2,3}$ are its roots, usually, it is chosen $e \equiv e_1+e_2+e_3 = 0$. As was indicated in \cite{EK:1993} the whole symmetry of (\ref{OPHam}) is the central and co-central extended loop group ${\tilde L}(SL(3))$. Note that the spectrum of quasi-periodic eigenfunctions is usually treated in the Bethe Anzatz formalism, see e.g. \cite{FV:1997}, \cite{NS:2009} and reference therein. It is worth mentioning that the spectrum was treated perturbatively in \cite{Perelomov et al.} and \cite{Langmann}.

If in (\ref{wp}) the trigonometric limit is taken, $\De \equiv g_2^3 + 27 g_3^3 = 0$, with one of the periods going to infinity, the Hamiltonian of $A_2$ trigonometric/hyperbolic Calogero-Moser-Sutherland model (3-body Sutherland model) occurs. If both invariants $g_2=g_3=0$ we arrive at the $A_2$-rational (or saying differently, at the 3-body Calogero-Moser) model. For future convenience we parameterize the invariants as follows
\begin{equation}
\label{wp-inv}
     g_2 = 12 (\ta^2 - \mu)\ ,\qquad  g_3=4 \ta (2\ta^2-3\mu)\ ,
\end{equation}
where $\ta, \mu$ are parameters.

The Hamiltonian (\ref{OPHam}) is translation-invariant, thus, it makes sense to introduce center-of-mass coordinates
\begin{equation}
\label{y}
    Y=\sum_1^3 x_i \ ,\ y_i = x_i - \frac{1}{3}Y\ ,
\end{equation}
with the condition $\sum_1^3 y_i=0$. Laplacian $\De^{(3)} \equiv \sum_{i=1}^3\frac{\pa^{2}}{\pa x_i^{2}}$ in these coordinates takes the form,
\[
    \De^{(3)}\ =\ 3\ \pa_Y^2\ +\ \frac{2}{3}\ \left(\frac{\pa^{2}}{\pa y_1^{2}}+\frac{\pa^{2}}{\pa y_2^{2}} - \frac{\pa^{2}}{\pa y_1 \pa y_2}\right)\ .
\]
Separating out center-of-mass coordinate $Y$ two-dimensional Hamiltonian arises
\begin{equation}
\label{OPHam2}
 {\cal H}_{A_2}\ =\ -\ \frac{1}{3}\ \left(\frac{\pa^{2}}{\pa y_1^{2}}+\frac{\pa^{2}}{\pa y_2^{2}} - \frac{\pa^{2}}{\pa y_1 \pa y_2}\right)\ +\
 \nu (\nu -1) \ \bigg(\wp (y_1-y_2)\ +\ \wp (2y_1+y_2) \ +\ \wp (y_1+2y_2)\bigg)\ .
\end{equation}

Since we will be interested by the general properties of the operator ${\cal H}_{A_2}$, without a loss of generality we can assume that the operator (\ref{OPHam2}) is defined on the real plane, $y_{1,2} \in {\bf R}^2$, while the fundamental domain of the Weierstrass function $\wp(x)$ is not fixed. The whole discrete symmetry of the Hamiltonian (\ref{OPHam2}) is $S^2 \oplus \mathbb{Z}_2 \oplus (T_r)^2 \oplus (T_c)^2$. It consists of permutation $S^2 (y_1 \lrar y_2)$, reflection $\mathbb{Z}_2 (y_{1,2} \lrar -y_{1,2})$ and four translations $T_{r,1(2)}:\ y_{1(2)} \rar y_{1(2)} + 1$ and $T_{c,1(2)}:\ y_{1(2)} \rar y_{1(2)} + i\ \ta_c$ (periodicity). Perhaps, $S^2 \oplus (T_r)^2 \oplus (T_c)^2$ can make sense as a double-affine $A_2$ Weyl group.

Let us consider a formal eigenvalue problem
\begin{equation}
\label{sp}
     {\cal H}_{A_2} \Psi\ =\ E \Psi\ ,
\end{equation}
without posing concrete boundary conditions. Assume $f(x)$ be the non-constant solution of the equation
\begin{equation}
\label{ff}
f'(x)^2=4 f(x)^3-12  \tau f(x)^2+12 \mu f(x)\ .
\end{equation}
Thus, it can be written as
\[
    f(x)\ =\ \wp (x | g_2, g_3) + \ta\ ,
\]
cf. (2),(3).
Now let us introduce the new variables
\begin{equation}
\label{trans}
x=\frac{f'(y_1)-f'(y_2)}{f(y_1) f'(y_2)-f(y_2) f'(y_1)}\, , \qquad
y=\frac{2(f(y_1)- f(y_2))}{f(y_1) f'(y_2)-f(y_2) f'(y_1)}\ ,
\end{equation}
which have the property
\[
  x(-y_1,-y_2)\ =\ x(y_1,y_2)\ ,\  y(-y_1,-y_2)\ =\ -y(y_1,y_2)\ .
\]
They are invariant with respect to the partial discrete symmetry of the Hamiltonian (\ref{OPHam2}): $S^2 \oplus (T_r)^2 \oplus (T_c)^2$. It can be shown that in the rational limit $\ta=\mu=0$, where the 3-body Calogero-Moser model emerges, the variables $x,y$ coincide with those found in R\"uhl-Turbiner \cite{RT:1995}

\renewcommand{\theequation}{8.{\arabic{equation}}}
\setcounter{equation}{0}
\begin{equation}
\label{e8.1}
        x =-( y_1^2+ y_2^2 + y_1 y_2),\qquad y = - y_1 y_2 (y_1 + y_2)\ .
\end{equation}
In the trigonometric limit $\mu=0$ the 3-body Sutherland Hamiltonian emerges in a form of the algebraic operator \cite{RT:1995}
\[
        x\  =\ \frac{1}{\al^2}\ [\cos(\al y_1)+\cos(\al y_2)+\cos(\al (y_1+y_2))-3]\ ,
\]
\begin{equation}
\label{e8.2}
    y\ =\ \frac{2}{\al^3}\ [\sin(\al y_1)+\sin(\al y_2)-\sin(\al (y_1 + y_2))]\ ,
\end{equation}
here $\al$ is a parameter such that $\ta = \al^2/12$\,.
\renewcommand{\theequation}{{\arabic{equation}}}
\setcounter{equation}{8}
It is worth noting that the variables (\ref{e8.1}), (\ref{e8.2}) being $A_2$ Weyl invariants were obtained making the averaging over some orbits in $A_2$ root space, see e.g. \cite{Turbiner:2013t}. It is an interesting open question whether (\ref{trans}) can be obtained as a result of averaging over some orbits, in particular, orbits generated by fundamental weights.

After tedious calculations it can be found that the $A_2$ elliptic Calogero-Moser potential (see (\ref{OPHam}), (\ref{OPHam2})) in the new variables (\ref{trans}) takes a rational form,
\begin{equation}
\label{potential}
  V(x,y)\ =\ \frac{3\nu(\nu-1)}{4}\ \frac{\Big(x+2\tau x^2+\mu x^3-6 (\mu-\tau^2) y^2 +
             3 \mu \ta x y^2\Big)^2}{D} \ ,
\end{equation}
where
\begin{equation}
\label{D}
    12 D(x,y)\ =\ 9 \mu^2 x^4 y^2 + 54 \ta \mu^2 x^2 y^4 + 27\mu^2(3\ta^2 - 4 \mu) y^6 -
    12 \mu x^5 - 72 \ta \mu x^3 y^2 -
\end{equation}
\[
108\mu (\ta^2  - 2\mu) x y^4 -
 12 \ta\,x^4 - 18 (4\ta^2 + 5\mu) x^2 y^2
 - 54 \ta(2\ta^2 - 3 \,\mu) y^4 - 4 x^3 - 108 \ta x y^2  - 27 y^2\ .
\]
It is worth noting that the potential (\ref{potential}) is symmetric in $y$, $V(x,y)=V(x,-y)$ as well as $D(x,y)=D(x,-y)$. Furthermore, the two-dimensional Laplacian in (\ref{OPHam2}) becomes the Laplace-Beltrami operator
\[
    \De_g(z_1,z_2)\ =\ g^{-1/2} \ \sum_{ij} \frac{\pa}{\pa z_i} g^{1/2} g^{ij} \frac{\pa}{\pa z_j}\ =\ g^{ij} \frac{\pa^2}{\pa z_i \pa z_j} +  \sum \frac{g^{ij}_{,i}}{2}\ \frac{\pa}{\pa z_j}\ ,
\]
where ${g^{ij}_{,i}} \equiv \frac{\pa g^{ij}}{\pa z_i}$,\
which in $(x,y)$-coordinates looks explicitly as
\[
\De_g(x,y;\ta,\mu) = 3\Big(\frac{x}{3}+\tau x^2+\mu x^3 +(\mu-\ta^2) y^2- \mu \ta x y^2 -
  \mu^2 x^2 y^2 \Big)\frac{\pa^2}{\pa x^2}\ +
\]
\begin{equation}
\label{LB-A2}
  y \Big(3 +8 \ta x+7 \mu x^2 - 3\mu \ta y^2 - 6 \mu^2 x y^2 \Big)\frac{\pa^2}{\pa x \pa y}\ +\
  \Big(-\frac{x^2}{3}+3\ta y^2+4\mu x y^2 -3\mu^2 y^4 \Big)\frac{\pa^2}{\pa y^2}\ +
\end{equation}
\[
  \Big(1+4 \ta x+5 \mu x^2-3\mu \ta y^2-6\mu^2 x y^2  \Big) \frac{\pa}{\pa x}\ +
   2 y \Big(2 \tau+ 3 \mu x-3\mu^2 y^2  \Big)\frac{\pa}{\pa y}\ .
\]
Thus, the flat contravariant metric, defined by the symbol of the Laplace-Beltrami operator in these coordinates, becomes two-parametric polynomial in $x,y$.  The Hamiltonian is the sum of the Laplace-Beltrami operator (\ref{LB-A2}) with polynomial coefficients and the rational potential (\ref{potential}).  Taking in the Laplace-Beltrami operator (\ref{LB-A2}), the rational limit $\ta=\mu=0,$ we arrive at the Laplace-Beltrami operator $\De^{(rat)}_g$ of the 3-body Calogero-Moser model \cite{RT:1995}. If  the trigonometric limit $\mu=0$ is taken, the Laplace-Beltrami operator $\De^{(trig)}_g$ of the 3-body Sutherland model emerges \cite{RT:1995}.

The denominator $D$ in (\ref{potential}) turns out to be equal to the determinant of the contravariant metric $D={\rm Det} (g^{ij}) = \frac{1}{g}$. It is worth noting some properties
of the determinant $D$: in the rational case, $D^{1/2}$ is the zero mode of the Laplace-Beltrami operator
\[
    \De^{(rat)}_g D^{1/2}\ =\ 0\ .
\]
In the trigonometric case
\[
    \De^{(trig)}_g D^{1/2}\ =\ -12 \ta D^{1/2}\ ,
\]
and in general case,
\[
    \De_g(x,y;\ta,\mu) D^{1/2}\ =\ -12 \ta\left(1- \mu(2x - 3\mu y^2)\right) D^{1/2}\ .
\]

It is easy to verify that the determinant $D(x,y)$ given by formula (\ref{D}) can be written as
\begin{equation}
\label{DW}
  D(x,y)\ =\ \frac{1}{12} W^2\ ,
\end{equation}
where the function
\begin{equation}
\label{W}
W = \frac{\pa y}{\pa y_2} \frac{\pa x}{\pa y_1}-\frac{\pa x}{\pa y_2} \frac{\pa y}{\pa y_1}\ ,
\end{equation}
is the Jacobian associated with the change of variables $(y_1, y_2) \rar (x,y)$\,.
The equation $w^2=12\, D(x,y)$ can be considered as the equation for the elliptic surface
\footnote{
In the case of $A_1$ elliptic model the variable $x$, which is the invariant with respect to the symmetry of the $A_1$ Hamiltonian $\mathbb{Z}_2 \oplus (T_r) \oplus (T_c)$, is equal to Weierstrass function,  $x=\wp(y)$ (see \cite{Lame:1989-2014}), the function $W=\frac{\pa x}{\pa y}$ is the Jacobian and the determinant $D(x)$ is a cubic polynomial $D=P_3 (x)$; the equation analogous to (\ref{DW}) defines the elliptic curve, $w^2 = P_3 (x)$.
}.
One can verify that the Jacobian $W$ admits a representation in factorized form,
\begin{equation}
\label{WF}
 W(y_1,y_2)=\frac{\si(y_1-y_2)\,\si(y_1+2 y_2)\,\si(y_2+2 y_1)}{\si_1^3(y_1)\, \si_1^3(y_2)\, \si_1^3(y_1+y_2)}\ .
\end{equation}
Here the Weierstrass $\si$-function \cite{WW:1927} has the parameters $g_i$ given by (\ref{wp-inv}) and $e=-\tau$ is a root of the  $\wp-$Weierstrass function, $\wp'(-\tau)=0$.
The function $\si_1$ is the $\si$-function associated with the half-period $\om$ corresponding to the root $-\tau$\,, thus, $\wp(\om)=-\ta$\,. By definition (see \cite{WW:1927}),
\[
 \si_1(x)=\frac{\si(x+\om)}{\si(\om)}\, \exp\Big(-\frac{\si'(\om)}{\si(\om)}\, x\Big)\ .
\]
Note that in the one-dimensional case $n=1$ the Jacobian becomes
\[
W(y_1)\ =\ -\wp'(y_1)\ =\ \frac{\si(2 y_1)}{\si_1(y_1)^4}\ ,
\]
see, \cite{WW:1927}, Ch.20, problem 24.

{\bf Conjecture.} For arbitrary $n$ the Jacobian
\[
W=\frac{\prod_{i>j}^{n+1}\si(y_i-y_j)}{\prod_{i=1}^{n+1} \si_1^{n+1}(y_i)}\ ,
\]
where $y_1+y_2+\cdots+y_{n+1}=0$.

There are two essentially different degenerations of the $\wp-$Weierstrass function into the trigonometric case:
(I) when $e=-\ta$ is double root, thus, $e=2 \ta$ is the simple root and then $\mu=0$, and (II) when $e=-\ta$ is a simple root and $\mu=\frac{3}{4} \ta^2$\,. In both cases
\[
\wp(x) \to \frac{\al^2}{4\sin^2{\frac{\al x}{2}}}-\frac{\al^2}{12}
\]
but in case (I) $\ta=\frac{\al^2}{12}$ whereas in case (II)
$\tau=-\frac{\al^2}{6}$\,. For the first degeneration the Jacobian is
\renewcommand{\theequation}{14.{\arabic{equation}}}
\setcounter{equation}{0}
\begin{equation}
\label{e14.1}
   W(y_1,y_2)\ =\ \frac{8}{\al^3}\ \sin \frac{\al(y_1-y_2)}{2}\,\sin \frac{\al(y_1+2 y_2)}{2}\,
   \sin \frac{\al(2 y_1+y_2)}{2}
\end{equation}
and for the second one the Jacobian is factorized as follows
\begin{equation}
\label{e14.2}
W(y_1,y_2)= \frac{8}{\al^3}\  \frac{\sin \frac{\al(y_1-y_2)}{2}\,\sin \frac{\al(y_1+2 y_2)}{2}\,\sin \frac{\al(2 y_1+y_2)}{2}}
  {\cos^3 \frac{\al y_1}{2} \cos^3 \frac{\al y_2}{2} \cos^3 \frac{\al(y_1+y_2)}{2}}\ .
\end{equation}
where $\al$ is a parameter such that $\ta = \al^2/12$\,.
\renewcommand{\theequation}{{\arabic{equation}}}
\setcounter{equation}{14}
The factorization of the case (I) cannot be generalized to the elliptic case where, in general, we have no multiple roots.

Surprisingly, the gauge rotation of (\ref{OPHam2}) with determinant $D$ (\ref{D}) as a gauge factor
\begin{equation}
\label{gs}
 h(x,y)\ =\ - 3 D^{-\frac{\nu}{2}}\,({\cal H}_{\rm A_2} - E_0)\,D^{\frac{\nu}{2}}\ ,
\end{equation}
where $E_0=3\nu(3\nu+1) \tau$,
transforms the Hamiltonian ${\cal H}_{\rm A_2} - E_0$ into the algebraic operator(!),
\[
h(x,y)\ =\ \Big(x+3 \tau x^2+3 \mu x^3 +3 (\mu-\ta^2) y^2-3 \mu \ta x y^2
  -3\mu^2 x^2 y^2 \Big)\frac{\pa^2}{\pa x^2}\ +
\]
\[
  y \Big(3+8\ta x+7\mu x^2-3\mu \ta y^2-6\mu^2 x y^2 \Big)\frac{\pa^2}{\pa x \pa y}\ +
\]
\begin{equation}
\label{A2}
  \frac{1}{3}\Big(-x^2+9 \tau y^2+12 \mu x y^2-9\mu^2 y^4  \Big)\frac{\pa^2}{\pa y^2}\ +
\end{equation}
\[
  (1+3\nu) \Big(1+4 \ta x+5 \mu x^2-3\mu \ta y^2-6\mu^2 x y^2  \Big) \frac{\pa}{\pa x}\ +
   2 (1+3\nu) y\Big(2 \tau+ 3 \mu x-3\mu^2 y^2  \Big)\frac{\pa}{\pa y}\ +
\]
\[
   3\nu(1+3\nu) \mu \Big(2 x-3\mu y^2 \Big)\ .
\]
Note the important $\mathbb{Z}_2$ symmetry property of this gauge-rotated Hamiltonian,
\[
h(x,y)=h(x,-y)\ .
\]
Thus, it follows that in the variables $(u=x, v=y^2)$ the operator $h$ remains algebraic,
\[
h(u,v)\ =\ \Big(u+3 \tau u^2+3 \mu u^3 +3 (\mu-\ta^2) v-3 \mu \ta u v
  -3\mu^2 u^2 v \Big)\frac{\pa^2}{\pa u^2}\ +
\]
\begin{equation}
\label{A2m}
2 v \Big(3+8\ta u+7\mu u^2-3\mu \ta v -6\mu^2 u v \Big)\frac{\pa^2}{\pa u \pa v}\ +
  4v \Big(-\frac{u^2}{3}+3 \tau v+4 \mu u v-3\mu^2 v^2  \Big)\frac{\pa^2}{\pa v^2}\ +
\end{equation}
\[
  (1+3\nu) \Big(1+4 \ta u+5 \mu u^2-3\mu \ta v-6\mu^2 u v  \Big) \frac{\pa}{\pa u}\ +\
\]
\[
   2\bigg(-\frac{u^2}{3} + \ta (7+12\nu) v + 2 \mu (5+9\nu) u v
    - 9 \mu^2 (1+2\nu) v^2  \Bigg)\frac{\pa}{\pa v}\ +
\]
\[
   3\nu(1+3\nu) \mu \Big(2 u -3\mu v \Big)\ .
\]
It is an alternative algebraic form of the gauge-rotated operator (\ref{gs}). Note that the variables $(u, v)$ are invariants with respect to the whole discrete symmetry of the Hamiltonian (\ref{OPHam2}): $S^2 \oplus \mathbb{Z}_2 \oplus (T_r)^2 \oplus (T_c)^2\ ,$ unlike the variables
$(x, y)$.

The operator $h(x,y)$ has also a property of self-similarity:
the gauge-rotated operator $\tilde h=D^{-m} h D^{m}$ with $m=(\frac{1}{2}-\nu)$  has polynomial coefficients as well as the corresponding gauge-rotated operator ${\tilde k}_{\rm A_2}=D^{-m} k_{\rm A_2} D^{m}$ (see below). It is easy to verify that $$\tilde h_{\nu}=h_{4-3\nu} - 12 (1 - 2 \nu) \tau\ .$$ Evidently, the operator $\tilde h_{\nu}$ has the same functional form of the potential (\ref{potential}) as for the operator $h_{\nu}\ .$

Let
\[
J_1=\frac{\pa}{\pa x}\ , \ J_2=\frac{\pa}{\pa y}\ , \ J_3=x \frac{\pa}{\pa x}\ ,\ J_4=y \frac{\pa}{\pa x}\ , \ J_5=x \frac{\pa}{\pa y}\ ,\ J_6=y \frac{\pa}{\pa y}\ ,
\]
\begin{equation}
\label{sl3}
 J_7=x (x \frac{\pa}{\pa x}+ y \frac{\pa}{\pa y}+3\nu)\quad , \quad   J_8=y (x \frac{\pa}{\pa x}+ y \frac{\pa}{\pa y}+3\nu)\ .
\end{equation}
Notice that these formulas define a representation $(-3\nu, 0)$ of the Lie algebra $sl(3)$ in differential operators of first order (see e.g. \cite{RT:1995}). If the spin (mark) of representation
$$-3\nu = n$$ takes an integer value, a finite-dimensional representation appears: the space of polynomials
\begin{equation}
\label{Pn}
   {\mathcal P}_n\ =\ <x^p y^q\ |\ 0 \leq p+q \leq n >\ ,\quad
   \dim {\mathcal P}_n = \frac{(n+2)(n+1)}{2}\ ,
\end{equation}
is preserved by $J$'s. It is worth mentioning that the space (\ref{Pn}) is invariant under
%(up to overall multiplicative factor) with respect to two-dimensional projective (M\"obius)
linear transformations,
\[
%   x \rar \frac{a_1 x + b_1 y +c_1}{a x + b y +c}\quad ,\quad
x \rar {a_1 x + b_1 y +c_1}\ ,\
%   y \rar \frac{a_2 x + b_2 y +c_2}{a x + b y +c}
y \rar {a_2 x + b_2 y +c_2}\ ,
\]
where $a,b,c$'s are parameters.
It can be easily shown by direct calculation that for any $\nu$ the operator $h$ (\ref{A2}) can be rewritten in terms of $sl(3)$ generators,
\begin{equation}
\label{Hsl3}
  h\ =\ (1+3\nu) J_1 J_3 - 3\nu J_3 J_1+3 J_1 J_6+3 \tau J_3^2+ 6 \tau(1-4 \nu) J_3 J_6\ +
  3(\mu-\tau^2) J_4^2\ +
\end{equation}
\[
  \tau (1 + 12 \nu) (J_4 J_5+J_5 J_4) +
  2 (1+3\nu) \mu J_3 J_7-3\mu \ta J_4 J_8-\frac{1}{3} J_5^2+ 3\tau J_6^2\ +
\]
\[
 4 \mu J_6 J_7+\mu (1 - 6 \nu) J_7 J_3 -3\mu^2 J_8^2 \ .
\]
Thus, the gauge-rotated Hamiltonian $h(x,y)$ describes an $sl(3)$-quantum Euler-Arnold top in a constant magnetic field. Hence, the 3-body elliptic Calogero-Moser model with an arbitrary coupling constant, is equivalent to $sl(3)$-quantum Euler-Arnold top in a constant magnetic field. If the coupling constant in (\ref{OPHam}) takes discrete values
\begin{equation}
\label{kappa}
    \ka \ =\ \frac{n}{9}\ (n+3)\ ,\ n=0,1,2,\ldots\ ,
\end{equation}
the Hamiltonian $h(x,y)$ as well as the Hamiltonian (\ref{OPHam2}) both have a finite-dimensional invariant subspace ${\mathcal P}_n$ . Hence, there may exist a finite number of analytic eigenfunctions of the form
\begin{equation}
\label{eigen}
 \Psi_{n,i}\ =\ P_{n,i}(x,y)\ \,D^{-\frac{n}{6}}\ , \quad i=1,\ldots ,\ \frac{(n+2)(n+1)}{2} \ ,
\end{equation}
where polynomial $P_{n,i}(x,y) \in {\mathcal P}_n$, see (\ref{Pn}). For example, for $n=0$ (at zero
coupling),
\[
    E_{0,1} = 0\ ,\ P_{0,1} = 1\ .
\]
For $n=1$ at coupling
\[
    \ka\ =\ \frac{4}{9}\ ,
\]
the operator $h$ has a three-dimensional kernel (three zero modes) of the type $(a_1 x+a_2y+b)$. The first non-trivial solutions appear for $n=2$ and
\[
    \ka\ =\ \frac{10}{9}\ .
\]
There exist six polynomial eigenstates.
Eigenvalues are given by the roots of the algebraic equation of degree 6,
\[
     (E^2+ 4 \ta E + 4 \mu)(E^2+ 8 \ta E + 4 \mu + 12 \ta^2)(E^2+ 12 \ta E + 4 \mu + 16 \ta^2)\
     =\ 0\ ,
\]
given by
\[
    E_{\pm}^{(1)} = - 2(\ta \pm  \sqrt {\ta^2 - \mu})\ ,\quad E_{\pm}^{(2)} = - 2(2\ta \pm \sqrt {\ta^2 - \mu})\ , \quad E_{\pm}^{(3)} = - 2(3\ta \pm \sqrt {5\ta^2 - \mu})\ .
\]
The corresponding eigenfunctions are of the form
$(a_1 x^2+a_2 xy + a_3 y^2 +b_1 x+b_2y+c)$.
Using formulas (\ref{trans}) and (\ref{gs}), one can construct the corresponding eigenfunctions for the original Hamiltonian (\ref{OPHam}) in an explicit form.

\noindent
{\it Observation I:} Let us construct the operator
\begin{equation}
\label{ipar-xy}
       i_{par}^{(n)}(x,y)\ =\ \prod_{j=0}^n ({\cal J}^0(n) + j)\ ,
\end{equation}
where
\[
{\cal J}^0(n) = x\frac{\pa}{\pa x}+y\frac{\pa}{\pa y} - n\ ,
\]
is the Euler-Cartan generator of the algebra $sl(3)$ (\ref{sl3}). It can be immediately seen that the algebraic operator $h(x,y)$ (\ref{A2}), at integer $n$, commutes with $i_{par}^{(n)}(x,y)$,
\[
  [h(x,y)\ ,\ i_{par}^{(n)}(x,y)]:\ {\mathcal P}_n \ \rar \ 0\ .
\]
Hence, $i_{par}^{(n)}(x,y)$ is the particular integral \cite{Turbiner:2013p} of the $A_2$ elliptic model (\ref{OPHam2}).

It is known (see \cite{Olshanetsky:1983}) that $A_2$ elliptic model is (completely)-integrable having a certain 3rd order differential operator $k_{\rm A_2}$ as the integral. Perhaps, the easiest way to find this integral, is to look for it in a form of an algebraic differential operator of the 3rd order, $[h(x,y), k_{\rm A_2}(x,y)]=0$. In the explicit form it is given by the following expression
\begin{equation}
\label{Ksl3diff}
k_{\rm A_2}(x,y) \ =\ -2\nu(1+3\nu)(2+3\nu)\,\mu\,y\,(2\ta + 3\mu x - 3\mu^2 y^2)
\end{equation}
%\end{multline*}
%%%%%%%%%%%%%%%%
%%%%%%%%%%%%%%%%3
\begin{multline*}
 +\frac{1}{3}\,(1+3\nu)(2+3\nu) y
(\mu + 8 \ta^2 +28\mu \ta x + 21\mu^2 x^2 - 9\mu^2\ta y^2 - 18\mu^3 x y^2)
\frac{\pa}{\pa x}
\end{multline*}
%%%%%%%%%%%%%%%%1
\begin{multline*}
-\frac{2}{9}\,(1+3\nu)(2+3\nu)\,
(1 +4\tau\,x + 6\mu\,x^2 - 24\mu\,\ta y^2 - 36\mu^{2} x {y}^{2} + 27\mu^3 y^4)
\frac{\pa}{\pa y}
\end{multline*}
%%%%%%%%%%%%%%%%1
\begin{multline*}
  + (2+3\nu) y \Big(3\,\ta + 4(2\ta^2+\mu)x + 17\mu \ta x^2 + 8\mu^2 x^3  \\
 + 3\mu({\tau}^{2} -2\mu) y^2 - 6\mu^{2}\ta x y^2 - 6{\mu}^3 x^2 y^2\Big)
\frac{\pa^2}{\pa x^2}
\end{multline*}
%%%%%%%%%%%%%%%%4
\begin{multline*}
-\frac{2}{3}\,(2+3\nu)
\Big(x+4\ta x^2 + 5\mu\,x^3 + 3(\mu - 4\ta^2) y^2 -27\mu^2 x^2 y^2 \\
 - 33\mu\,\ta x y^2
+ 9\mu^2 \ta y^4 + 18\mu^3 x y^4 \Big)
\frac{\pa^2}{\pa x \pa y}
\end{multline*}
%%%%%%%%%%%%%%%%5
\begin{multline*}
 - (2+3\nu)y(1+\frac{8}{3}\tau\,x+3\mu\,x^2 -7\mu \ta y^2 - 10\mu^2 x y^2 + 6\mu^3 y^4)
\frac{\pa^2}{\pa y^2}
\end{multline*}
%%%%%%%%%%%%%%%%6
\begin{multline*}
 +y \Big(1 + 5\ta x + 2(2\mu  +3\ta^2) x^2 + 3\mu (\ta^2 - 2\mu) x y^2 + 9 \mu \ta x^3
\\
 - \ta(3\mu - 2\ta^2) y^2 + 3\mu^{2} x^4 - 3\mu^{2}\ta x^2 y^2 - 2\mu^3 x^3 y^2 \Big)
\frac{\pa^3}{\pa x^3}
\end{multline*}
%%%%%%%%%%%%%%%%8
\begin{multline*}
+\Big(
 -\frac{2}{3} x^2+ 2(5{\tau}^{2}+\mu) x y^2 - 2\ta x^3 + 3\ta y^2-2\mu\,x^4 +
 3\mu(\ta^{2}-2\mu) y^4 + 19\mu\,\tau x^2 y^2
\\
 -6{\mu}^3 x^2 y^4 + 10\mu^{2} x^3 y^2 -6 \mu^2 \ta x y^4 \Big)
\frac{\pa^3}{\pa x^2 \pa y}
\end{multline*}
%%%%%%%%%%%%%%%%9
\begin{multline*}
- y \Big( x + \frac{10}{3} \tau\,{x}^{2} +
 \frac{11}{3}\,\mu\,{x}^{3} - 13\mu\,\ta x y^2 + 3(\mu-2\ta^2) y^2
  -11\mu^2 x^2 y^2 \\
  + 3\mu^2\ta y^4 + 6\mu^3 x y^4 \Big)
\frac{\pa^3}{\pa x \pa y^2}
\end{multline*}
%%%%%%%%%%%%%%%%10
\begin{multline*}
- \Big(y^2 + \frac{2}{27} x^3 + 2\tau\,x y^2 - 3\mu\,\ta y^4 +
 \frac{5}{3}\mu\,x^2 y^2 - 4{\mu}^{2} x y^4 + 2{\mu}^{3}y^6 \Big)
\frac{\pa^3}{\pa y^3}\ .
\end{multline*}
%%%%%%%%%%%%%%%%7
This operator is invariant with respect to $y \rar -y$,
\[
k_{\rm A_2}(x,y)\ =\ k_{\rm A_2}(x,-y)\ ,
\]
similarly to the gauge rotated Hamiltonian $h(x,y)$ (see (\ref{A2})). Thus, after the change of variables $(x, y) \rar (u=x, v=y^2)$ the operator $k_{\rm A_2}(u,v)$ remains algebraic.
Let us note for  $(2+3\nu)=0$ or, saying differently, for $n=2$ the operator $k_{\rm A_2}$ becomes a 3rd order homogeneous differential operator, it contains 3rd derivatives only. This operator can be rewritten in terms of $sl(3)$-generators,
\begin{equation}
\label{Ksl3}
 k_{\rm A_2}\ =\ J_1^2 J_4 +3 (2+3\nu)\tau J_1 J_3 J_4 -\frac{2}{9}  (1+3\nu)(2+3\nu) J_1 J_3 J_5 +\
\end{equation}
\[
 3 \tau J_1 J_4 J_6+\nu (2+3\nu) J_1 J_5 J_3 - 3\nu J_1 J_6 J_5 - (1+9 \nu) \tau J_3 J_1 J_4\ +
\]
\[
  \frac{1}{3}\big(12 \mu  +12 \tau^2 - (1+3\nu)(11\mu+16 \tau^2) + (1+3\nu)^2 (\mu+8 \tau^2)\big) J_3^2 J_4-\frac{8}{9} (1+3\nu)(2+3\nu) \tau J_3^3 J_5\ +
\]
\[
  4 (2+3\nu)(1 - 3\nu) \mu \ta J_3^2 J_8\ +
\]
\[
  \frac{2}{3}\big(3 \tau^2+(1+3\nu)(5\mu+4\tau^2)-(1+3\nu)^2 (\mu+8 \tau^2)\big) J_3 J_4 J_3 +
  \big(\mu + 8 \tau^2 +2 (1+3\nu)(\mu - 4\tau^2)\big) J_3 J_4 J_6\ +
\]
\[
  \frac{2}{9}(1+36\nu+72\nu^2) \tau J_3 J_5 J_3-(1-3\nu) J_3 J_6 J_2 -
    \frac{4}{3}(1+6\nu) \tau J_3 J_6 J_5+2(2+3\nu) \mu^2 J_3 J_7 J_8\ +
\]
\[
  -4(1+3\nu)\mu \ta J_3 J_8 J_6+\frac{1}{3}(1+3\nu)(2+3\nu)(\mu+8\tau^2) J_4 J_3^2 -
%\]
%\[
  (\mu(1+6\nu)- 2(5+12\nu)\ta^2)J_4 J_3 J_6\ -
\]
\[
 \frac{4}{3}(1+3\nu)(2+3\nu)\mu \ta J_4 J_3 J_7-\ta (3 \mu - 2 \ta^2) J_4^3 -
 3 \mu (2 \mu-\ta^2) J_4^2 J_8\ -\ 3 (\mu-2 \tau^2) J_4 J_6^2\ +
\]
\[
  2(7+6\nu)\mu \ta J_4 J_6 J_7 - 3 \mu^2 \ta J_4 J_8^2\ -\ \frac{1}{9}(2+9\nu^2) J_5 J_3 J_1-\frac{4}{9}(1+18\nu^2) \tau J_5 J_3^2\ -
\]
\[
  \frac{4}{3}(2+3\nu)\mu J_5 J_3 J_7 - \frac{2}{27} J_5^3\ +
 \frac{2}{3}(1+6\nu)\mu J_5 J_7 J_3 - J_6 J_2 J_6 -
 2(1-4\nu) \tau J_6 J_5 J_3 \ -
\]
\[
 - 2 \tau J_6 J_5 J_6 - \frac{5}{3} \mu J_6 J_5 J_7\ -
 \frac{1}{3}\mu \ta\big(5 - 72\nu^2 \big) J_7 J_3 J_4\ -\
 \mu^2 (1+6\nu)\mu^2\big) J_7 J_3 J_8 +
\]
\[
 4\mu^2 J_7 J_8 J_6 + 12\mu \ta J_8 J_6^2 - 9 \mu \ta J_6 J_8 J_6 - 2 \mu^3 J_8^3\ .
\]
It is evident that if $-3\nu=n$ the operator (\ref{Ksl3diff}) has the space ${\mathcal P}_n$ as a finite-dimensional invariant subspace. It seems natural to assume that the gauge-rotated integral
$k_{\rm A_2}$ written in variables $x_1,x_2,x_3$,
\[
    K_{\rm A_2}\ =\ D^{\frac{\nu}{2}}\,k_{\rm A_2}\,D^{-\frac{\nu}{2}}\ ,
\]
should coincide with the integral found recently by Oshima \cite{Oshima:2007}.

An important observation should be made about a connection of the determinant (\ref{D}) $D \equiv D(\ta, \mu)$ with discriminants. It can be shown that $D$ being written in Cartesian coordinates has the factorized form,
\[
   D(0,0)\ =\ 4 x^3 + 27 y^2 \ \sim \ (y_1-y_2)^2 (y_1-y_3)^2 (y_2-y_3)^2\ ,
\]
so, it is the discriminant of the cubic equation;
\[
  D(\ta,0)\ = \ 12\tau\,x^4 + 4 x^3 + 72{\tau}^2 x^2 y^2 + 108\tau x y^2 + 27 y^2 + 108 {\tau}^3 y^4\ \sim
\]
\begin{equation}
\label{Dtrig}
 \sin^2 \al(y_1-y_2) \sin^2 \al(y_1-y_3)  \sin^2 \al(y_2-y_3) \ ,
\end{equation}
is a trigonometric discriminant, where $\ta=\frac{\al^2}{3}$. In general,
$D(\ta, \mu)=\frac{W^2(\ta, \mu)}{12}\,,$ where (cf. (\ref{WF}))
\begin{equation}
\label{Dell}
W(\ta, \mu)\ \sim \frac{\si(y_1-y_2)\,\si(y_2-y_3)\,\si(y_3-y_1)}{\si_1^3(y_1)\, \si_1^3(y_2)\, \si_1^3(y_3)} \ ,
\end{equation}
and  $\si(x)$ and  $\si_1(x)$ are the Weierstrass $\si$ functions (see \cite{WW:1927}), might be an elliptic discriminant.

It also must be noted that the operator $h(u,v)$ (see (\ref{A2m})) can be rewritten in terms of the generators of the algebra $g^{(2)}$: the infinite-dimensional, eleven generated algebra of differential operators introduced in \cite{Turbiner:1998} (see for a discussion \cite{Turbiner:2013t}). It is spanned by the Euler-Cartan generator
\begin{equation}
\label{J0-g2}
{\tilde {\cal J}}_0(n)\ =\ u\pa_u \  +\ 2 v\pa_v \ - \ n\ ,
\end{equation}
(cf. with the Euler-Cartan generator ${\cal J}_0(n)$ of $sl(3)$-algebra), and
\[
 {\cal J}^1\  =\  \pa_u\ ,\
 {\cal J}^2_n\  =\ u \pa_u\ -\ \frac{n}{3} \ ,\
 {\cal J}^3_n\  =\ 2 v\pa_v\ -\ \frac{n}{3}\ ,
\]
\begin{equation}
\label{gl2r}
       {\cal J}^4_n\  =\ u^2 \pa_u \  +\ 2 u v \pa_v \ - \ n u\ =\ u {\tilde {\cal J}}_0(n)\ ,
\end{equation}
\[
 {\cal R}_{0}\  = \ \pa_v\ ,\ {\cal R}_{1}\  = \ u\pa_v\ ,\ {\cal R}_{2}\  = \ u^{2}\pa_v\ ,\
\]
\[
 {\cal T}_0\ =\ v\pa_{u}^2 \ ,\ {\cal T}_1\ =\ v\pa_{t} {\tilde {\cal J}}_0{(n)}\ ,\
 {\cal T}_2\ =\
 v {\tilde {\cal J}}_0{(n)}\ ({\tilde {\cal J}}_0{(n)} + 1) = \ v {\tilde {\cal J}}_0{(n)}\ {\tilde {\cal J}}_0{(n-1)}\ ,
\]
where $n$ is a parameter. If $n$ takes an integer value, the algebra $g^{(2)}$ has a common invariant subspace (finite-dimensional representation space),
\begin{equation}
\label{Qn}
   {\mathcal Q}_n\ =\ <u^p v^q\ |\ 0 \leq p+2q \leq n >\ ,
\end{equation}
where it acts irreducibly. The space (\ref{Qn}) is invariant with respect to the
%weighted projective
polynomial transformations
\[
     u \rar u + A_u \quad ,\quad
     v \rar v + A_v u^2 + B u + C \ ,
\]
where $A_{u,v}, B, C$ are constants.

Note that if in (\ref{kappa}) the parameter $n$ is an integer and thus, $-3\nu=n$,  the operator (\ref{A2m}) at $\nu=-n/3$ has the finite-dimensional invariant subspace (\ref{Qn}). This operator can be rewritten in terms of generators (\ref{gl2r}). Hence, the algebra $g^{(2)}$ is a hidden algebra of the $A_2$ elliptic Calogero-Moser model as well being alternative to the hidden algebra $sl(3)$. Let us note that it was already known that the algebra $g^{(2)}$ is the hidden algebra of the $G_2$ rational and trigonometric models \cite{Turbiner:1998}. Below we will show that it remains the hidden algebra of the $G_2$ elliptic model!
Note that can be shown that the square of operator (\ref{Ksl3diff}), $k_{\rm A_2}^2$ written in the variables $(u,v)$ is the algebraic operator and it commutes with the Hamiltonian (\ref{A2m}), $[h(u,v), k_{\rm A_2}^2(u,v)]=0$.

It turns out that if we add to the operator $h(u,v)$ (\ref{A2m}) the operator
\begin{equation}
\label{G2add}
h_m(u,v)\ =\ 6 (1+2 \tau u+\mu u^2) \frac{\pa}{\pa u}\ +
4 (-u^2+3 \tau v+3 \mu u v)\frac{\pa}{\pa v}\ +18\, \nu \mu \,u
\end{equation}
\[
   =\ 6 {\cal J}^1 - 4 {\cal R}_{2} + 6 \ta {\cal J}^2_{-3\nu} + 6 \ta {\cal J}^3_{-3\nu} + 6 \mu {\cal J}^4_{-3\nu} - 12 \ta \nu\ ,
\]
the resulting operator
\begin{equation}
\label{hG2}
 h_{G_2}(u,v)\ =\ h(u,v) + \la h_m(u,v)\ ,
\end{equation}
where $\la$ is an arbitrary parameter, is an algebraic form for the $G_2$ elliptic model.
Let us denote by $\tilde D(u,v)$ the right hand side of (\ref{D}) written in the variables
$u=x,\, v=y^2$\ . Then the gauge transformation
\[
\tilde h_{G_2}(u,v)=p\, \bigg(h_{G_2}(u,v) + 3\nu (3 \nu + 6\la + 1)\ta \bigg)\, p^{-1}\ ,
\]
where $p=v^{\frac{3\la}{2}} {\tilde D}^{\frac{\nu-\la}{2}}$\ , brings the operator $h_{G_2}$ to the Schroedinger operator form
\[
 {\tilde h}_{G_2}\ =\ \De_g+ \la(3\la-1)\frac{u^2}{v}+9 (\nu-\la)(\nu-\la-1)\frac{(u+2 \tau u^2+\mu u^3-6 \mu v+6 \tau^2 v+3 \mu \tau\, u v)^2}{\bar D}\ .
\]
% \[
% -(n_2+3 n_1)(3 n_1+3 n_2+1) \ta \ .
% \]
Now the transformation
%\begin{equation}
%\label{trans-uv}
\[
   u=\frac{f'(y_1)-f'(y_2)}{f(y_1) f'(y_2)-f(y_2) f'(y_1)}\, , \qquad
   v=\Big(\frac{2(f(y_1)- f(y_2))}{f(y_1) f'(y_2)-f(y_2) f'(y_1)}\Big)^2\ ,
\]
%\end{equation}
where $f$ is defined by (\ref{ff}),  relates the Hamiltonian ${\tilde h}_{G_2}$ with the Hamiltonian of $G_2$ elliptic model \cite{Olshanetsky:1983}
\[
 {\cal H}_{G_2}\ =\ -\ \frac{1}{3}\ \left(\frac{\pa^{2}}{\pa y_1^{2}}+
           \frac{\pa^{2}}{\pa y_2^{2}} - \frac{\pa^{2}}{\pa y_1 \pa y_2}\right)\ +\
(\nu-\la)(\nu-\la-1) \ \Big(\wp (y_1-y_2)\ +\ \wp (2y_1+y_2) \ +\ \wp (y_1+2y_2)\Big)
\]
\begin{equation}
\label{OPHam2G}
 +\ \la(3\la-1) \ \Big(\wp (y_1)\ +\ \wp (y_2) \ +\ \wp (y_1+y_2)\Big)\ .
% -(n_2+3 n_1)(3 n_1+3 n_2+1) \ta\ .
\end{equation}
The $G_2$ elliptic Hamiltonian is characterized by two coupling constants which can be parameterized as $\ka = (\nu-\la)(\nu-\la-1)$ (see (\ref{OPHam})) and $\ka_2 = \la(3\la-1)$.
If $\ka_2=0$, the $A_2$ elliptic model occurs. If $\nu = -\frac{n}{3}\ ,\ n=0,1,2,\ldots$ the $G_2$ elliptic Hamiltonian
\[
 {\cal H}_{G_2}\ =\ -\ \frac{1}{3}\ \left(\frac{\pa^{2}}{\pa y_1^{2}}+
           \frac{\pa^{2}}{\pa y_2^{2}} - \frac{\pa^{2}}{\pa y_1 \pa y_2}\right)\ +\
(\frac{n}{3}+\la)(\frac{n}{3}+\la+1) \ \Big(\wp (y_1-y_2)\ +\ \wp (2y_1+y_2) \ +\ \wp (y_1+2y_2)\Big)
\]
\begin{equation}
\label{OPHam2G-int}
 +\ \la(3\la-1) \ \Big(\wp (y_1)\ +\ \wp (y_2) \ +\ \wp (y_1+y_2)\Big)\ ,
% -(n_2+3 n_1)(3 n_1+3 n_2+1) \ta\ .
\end{equation}
has a number of polynomial eigenfunctions. They have the form
\begin{equation}
\label{eigenG2}
 \Psi_{n,i}\ =\ Q_{n,i}(u,v)\, v^{\frac{3\la}{2}} \,{\tilde D}^{\frac{\nu-\la}{2}}\ , \quad i=1,\ldots ,\ \dim {{\mathcal Q}_n} \ ,
\end{equation}
where polynomial $Q_{n,i}(u,v) \in {\mathcal Q}_n$, see (\ref{Qn}). If $\la=\frac{1}{3}$, the coupling constant $\ka_2=0$, the $G_2$ elliptic Hamiltonian degenerates to $A_2$ elliptic Hamiltonian (\ref{OPHam2}) which has polynomial eigenfunctions
\begin{equation}
\label{eigenG2-A2}
 \Psi_{n,i}\ =\ Q_{n,i}(u,v)\, v^{\frac{1}{2}} \,{\tilde D}^{-\frac{n+1}{6}}\ , \quad i=1,\ldots ,\ \dim {{\mathcal Q}_n} \ ,
\end{equation}
(c.f.(\ref{eigenG2})) at coupling constant,
\begin{equation}
\label{kappaG2}
    \ka \ =\ \frac{n+1}{9}\ (n+4)\ ,\ n=0,1,2,\ldots\ ,
\end{equation}
(c.f.(\ref{kappa})).

It is known that the $G_2$ elliptic model has the integral in the form of a 6th order differential operator (see \cite{Olshanetsky:1983}). It can be easily shown that there must exist a differential operator $k_m(u,v)$ of degree less than six such that
\[
   k_{\rm G_2}\ =\ k_{\rm A_2}^2(u,v) + \la k_m(u,v)\ ,
\]
commutes with the $G_2$ elliptic Hamiltonian (\ref{hG2}). It is evident that $k_0(u,v)$ can be rewritten in terms of the generators of the algebra $g^{(2)}$. This will be calculated elsewhere.

\noindent
{\it Observation II:} Let us construct the operator
\begin{equation}
\label{ipar-uv}
       i_{par}^{(n)}(u,v)\ =\ \prod_{j=0}^n ({\tilde {\cal J}}^0(n) + j)\ ,
\end{equation}
where
\[
{\tilde {\cal J}}^0(n) = u\frac{\pa}{\pa x} + 2 v \frac{\pa}{\pa y} - n\ ,
\]
is the Euler-Cartan generator of the algebra $g^{(2)}$ (\ref{J0-g2}). It can be immediately seen that the algebraic operator $h(u,v)$ (\ref{hG2}) at integer $n$ commutes with $i_{par}^{(n)}(u,v)$,
\[
  [h(u,v)\ ,\ i_{par}^{(n)}(u,v)]:\ {\mathcal Q}_n \ \rar \ 0\ .
\]
Hence, $i_{par}^{(n)}(u,v)$ is the particular integral \cite{Turbiner:2013p} of the $G_2$ elliptic model (\ref{OPHam2}).

In this paper we demonstrate that both $A_2$ and $G_2$ elliptic models belong to two-dimensional quasi-exactly-solvable (QES) problems \cite{Turbiner:1988, Turbiner:1994}. We show the existence of an algebraic form of the $A_2$ elliptic Hamiltonian, which is the second order polynomial element of the universal enveloping algebra $U_{sl(3)}$ and an algebraic form of the $G_2$ elliptic Hamiltonian, which is the element of the algebra $g^{(2)}$. We construct explicitly the integral for the $A_2$ case - commuting with the Hamiltonian - as the third order polynomial element of the universal enveloping algebra $U_{sl(3)}$. If the algebra $U_{sl(3)}$ appears in a finite-dimensional representation, those elements possess a finite-dimensional invariant subspace. This phenomenon happens for a discrete sequence of coupling constants (\ref{kappa}) for which both polynomial eigenfunctions and a particular integral occur. In a similar way, if the algebra $g^{(2)}$ appears in a finite-dimensional representation those elements possess a finite-dimensional invariant subspace. This also happens for $G_2$ elliptic model: for one-parametric family of coupling constants, a number of polynomial eigenfunctions occur as well as a particular integral.

The situation looks very similar to the case of the $A_1$ elliptic model (the Lame Hamiltonian, see e.g. \cite{Lame:1989-2014}), where the new variable which transforms the $A_1$ elliptic Hamiltonian to the algebraic operator is $x = \frac{1}{\wp(y_1)}$. A generalization to $A_n$ elliptic models for $n > 2$ seems straightforward. It is worth noting that a certain algebraic form for a general $BC_n$ elliptic model was found some time ago in \cite{DGU:2001, Brihaye:2003} (see also \cite{Lame:1989-2014}). The existence of the $sl(n)$ hidden algebra structure was also shown, that is equivalent to $sl(n)$ quantum Euler-Arnold top. Such generalizations can be regarded as a multivariate generalization of the Lame Hamiltonian as well as the Lame polynomials.

{\it Note added.}
After the present study was completed, based on the transformation (\ref{trans}), the following has been formulated

\noindent
{\it Conjecture (M. Matushko, August 2014)}. The analog of transformation (\ref{trans}) for arbitrary $n$ is given by the solution of the linear system
\[
M {\bf u}={\bf e},
\]
where ${\bf u}=(u_1, \ldots ,u_n)^t$,  ${\bf e}=(1, 1, \ldots, 1)^t$ with
\[
M_j^i=\frac{d^{j-1} \wp(y_i)}{d y_i^{j-1}}\ .
\]
%We are going to devote a separate paper on the subject.

It is evidently correct for $n=1$. Validity of this conjecture will be checked elsewhere.
It is worth commenting that the determinant of this linear system is the elliptic generalization
of the Van der Monde determinant (see e.g. \cite{WW:1927}). The vector ${\bf e}$ looks like the highest root among $A_n$ roots in the basis of simple roots.

\begin{acknowledgments}
  The authors are grateful to Maxim Kontsevich for the interest to the present work and numerous valuable discussions, V.V.S. thanks Masha Matushko for interest to the work and useful discussions, A.V.T. is grateful to Sergei P Novikov for numerous important discussions on the subject and encouragement, and he thanks Alberto Grunbaum for an important remark as well as Harry Braden for the indication to a reference. We thanks anonymous referees for inspiring comments which improved our presentation. From the side of A.V.T. his work was supported in part by the University Program FENOMEC, and by the PAPIIT grant {\bf IN109512} and CONACyT grant {\bf 166189}~(Mexico).
\end{acknowledgments}

\end{document}